\documentclass[12pt]{article}
\usepackage{graphicx}
\begin{document}
\vspace*{3cm}
\begin{center}
{\Large Cherenkov-like shock waves associated with surpassing\\
the light velocity barrier}\\[4mm]
{G.N. Afanasiev\footnote{Corresponding author:
Phone:   +7 (09621) 63179\hspace*{5mm}Fax: +7 (09621) 65084,\\
\hspace*{43mm}E-mail: afanasev@thsun1.jinr.ru}
\ and V.G. Kartavenko}\\[6mm]
{Bogoliubov Laboratory of Theoretical Physics,\\
Joint Institute for Nuclear Research,\\
 Dubna, Moscow District, 141980, Russia}
\end{center}

%\author

\vspace*{1cm}
%\author
\vfill{}
\eject
\begin{abstract}
The effects arising from accelerated and decelerated motion of a point
charge inside  a medium are studied. The motion is manifestly
relativistic and may be produced by a constant uniform electric field.
It is shown  that in addition to the bremsstrahlung and Cherenkov shock
waves, the electromagnetic shock wave arises when the charge
particle velocity coincides with the light velocity in the medium.
For the accelerated motion this shock wave forming
an indivisible entity with the Cherenkov shock wave arrives
after the arrival of the bremsstrahlung shock wave.
For the decelerated motion the above shock wave
detaches from the charge at the moment when its velocity coincides with
the light velocity in the medium. This wave existing even after
termination of the charge motion of the charge propagates with the light
velocity in the medium.
It has the same singularity as the Cherenkov shock and
is more singular than the bremsstrahlung shock wave.
The space-time regions, where these shock waves exist, and conditions
under which they can be observed are determined.
\end{abstract}
\eject
\section{Introduction}
Although the Vavilov-Cherenkov effect is a well-established phenomenon
widely used in physics and technology (see. e.g., Frank's book [1]),
a lot of its aspects remain uninvestigated up to now.
In particular, it is not clear how a transition from
the sub-light velocity regime to the the super-light one takes place.
Some time ago (Tyapkin [2], Zrelov et al. [3]) it was suggested
that
alongside  with the usual Cherenkov and bremsstrahlung ($BS$)
shock waves, there should exist a shock wave associated with the
charged particle passing  the medium  light velocity $c_n$.
The consideration presented there was pure qualitative without
any formulae and numerical results.
It was based on the analogy with the phenomena occurring in acoustics
and hydrodynamics. It seems to us that this analogy is not complete and,
therefore, it cannot be  considered as a final proof.

Usually, treating the Vavilov-Cherenkov effect, one considers the charge
motion in an infinite medium with a constant velocity $v>c_n$. In the
absence of $\omega$ dispersion, there is no electromagnetic field ($EMF$)
before the Cherenkov cone accompanying the charge,
an infinite $EMF$ on the Cherenkov cone itself and finite values
of the $EMF$ strengths behind it ([1, 4, 5]). In this case,
information concerning the transition effects arising
when  charge velocity coincides with $c_n$ is lost
(except for the existence of the Cherenkov shock wave ($CSW$) itself).

The accelerated motion of the point charge in a vacuum was first considered
by Schott ([6]). Yet, his  qualitative consideration was pure  geometric,
not allowing  numerical investigations.

Tamm ([7]) approximately solved the following problem:
A point charge rests at a fixed  point of medium up to some moment
$t=-t_0$ after which it exhibits the instant infinite acceleration
and moves uniformly with the velocity greater
than the light velocity in the medium. At the moment $t=t_0$ the charge
decelerates instantly  and rests afterwards. Later, this problem
was  numerically investigated by Ruzicka and Zrelov ([8,9]). The
analytic solution of this problem in the absence of dispersion has been
found in [5]. However, in all these studies the
information concerning the transition effects arising when charge velocity
surpasses the light velocity in medium was lost (due to the instant charge
acceleration).

Another  possibility is to use smooth accelerated and decelerated charge
motion as a tool for the studying the above-mentioned transition effects.
Previously, the straight-line motion of a point charge with  constant
acceleration ($z=at^2$) has been considered  in [10].
This motion law is obtained from the relativistic equation
$$m\frac{d}{dt}\frac{v}{\sqrt{1-\beta^2}}=eE,\quad
v=\frac{dz}{dt},\quad \beta=v/c,$$
(m is the rest mass) for the following electric field directed
along the $z$ axis
$$E_z=\frac{2ma}{e(1-4az/c^2)^{3/2}}.\eqno(1.1)$$

For the case of accelerated motion it has been found there that two shock
waves arise when charge velocity coincides with $c_n$.
The first of them is the well-known
Cherenkov shock wave $C_M$ having the form of the finite
Cherenkov cone and propagating with the velocity of the charge.
The second of these waves ($C_L$), closing the Cherenkov cone and
propagating with the velocity $c_n$, is just the shock wave the existence
of which was qualitatively predicted by Tyapkin ([2]) and
Zrelov et al.([3]). These two waves form an indivisible
entity. As time goes, the dimensions of this complex grow,
but its form remains essentially the same.
The singularities of the $C_L$ and $C_M$ shock waves are
the same and much stronger than the singularity of the $BS$ shock wave
arising from the beginning of the charge motion.

For the case of decelerated motion it has been found in the same reference
[10] that an additional shock wave arises
at the moment when the charge velocity coincides with $c_n$.
This wave being detached from the charge exists even after termination
of its motion. It propagates with the velocity $c_n$ and has the same
singularity as  $CSW$.

The drawback of this consideration  is that the electric field (1.1)
maintaining charge motion tends to $\infty$ as $z$ approaches $c^2/4a$.
This singularity makes the creation of electric field (1.1)  be rather
problematic. This, in turn, complicates the experimental verification of
the shock waves mentioned above.

Here, we consider the straight-line motion of a point charge in a constant
uniform electric field (which is much easier to create than the singular
electric field (1.1)) and evaluate  $EMF$ arising  from such a motion.
The arising motion law is manifestly relativistic. We suggest this motion
law to be given, disregarding the energy losses and the medium
influence on a moving charge.

Qualitatively, we confirm the results obtained  in [10] concerning the
existence of  the shock waves associated with the surpassing the light medium
velocity.

In the present approach, we take the refractive index to be independent of
$\omega$. This permits us to solve the problem under consideration
explicitly. The price for disregarding of the $\omega$ dependence is
the divergence of integrals quadratic in the Fourier transforms of
field strengths (such as the total energy flux). \\
This consideration is on the same footing as the first Tamm and
Frank papers on the Vavilov-Cherenkov effect in which the dispersion
and the influence of energy losses on the uniform point charge motion
were not taken into account. In spite of this, these papers correctly
predicted the location of the Vavilov-Cherenkov singularity.
The subsequent inclusion
of dispersion only slightly changed these results.\\
Another argument for the simplified treatment of the charge accelerated
motion (i.e., without $\omega$ dispersion) is due to Refs. [11]
where the uniform motion of a charge in medium with
a standard $\omega$ dependence of electric permittivity
($\epsilon(\omega)=1+\omega_L^2/(\omega_0^2-\omega^2+ip\omega)$)
was considered. It was shown there that such a $\omega$ dependence of
$\epsilon$ removes singularities of the field strengths and leads to
the appearance of many  maxima  of the radiated energy flux
behind the moving charge.
However, the main radiation maximum is at the same position as
in the absence of $\omega $ dispersion. Further, despite
the $\omega$ dispersion, the critical charge velocity (independent of
frequency and dependent on medium properties) exists below and above of
which the radiation spectrum differs drastically.
It turns out that for the
uniform charge motion the main features of the Cherenkov-Vavilov radiation
are the same with and without dispersion. Thus, we hope the same is true
for the $EMF$ radiated by the accelerated charge moving in medium.

\section{Statement of the physical problem}
Let a point charge  move inside the medium with the polarizabilities
$\epsilon$ and $\mu$ along the given trajectory $\vec\xi(t)$. Then, its
$EMF$ at the observation point ($\rho,z$) is given by the Lienard-Wiechert
potentials (see, e.g.,[12])
$$\Phi(\vec r,t)=\frac{e}{\epsilon}\sum_i\frac{1}{|R_i|},\quad
\vec A(\vec r,t)=\frac{e\mu}{c}\sum_i\frac{\vec v_i}{|R_i|},\quad
div\vec A+\frac{\epsilon\mu}{c}\dot\Phi=0\eqno(2.1)$$
Here
$$\vec v_i=(\frac{d\vec \xi}{dt})|_{t=t_i},\quad
R_i=|\vec r-\vec\xi(t_i)|-\vec v_i(\vec r-\vec\xi(t_i))/c_n$$
and $c_n$ is the light velocity inside the medium
($c_n=c/\sqrt{\epsilon\mu}$).
The summing in (2.1) is performed over all physical roots of the equation
$$c_n(t-t')=|\vec r-\vec\xi(t')|\eqno(2.2)$$
To preserve the causality, the time of radiation $t'$ should be smaller
than the observation time $t$.
Obviously, $t'$ depends on the coordinates $\vec r,t$ of the
observation point $P$.
With the account of (2.2) one gets for $R_i$
$$R_i=c_n(t-t_i)-\vec v_i(\vec r-\vec \xi (t_i))/c_n\eqno(2.3)$$
Consider the motion of the charged point-like particle
of the rest mass $m$ inside the medium according to the motion law ([12])
$$z(t)=\sqrt{z_0^2+c^2t^2}+C. $$
It may be realized in a constant electric field $E$ directed along
the $Z$ axis: $ z_0=|mc^2/eE|>0$. Here $C$ is an arbitraty constant.
We choose it from the condition $z(t)=0$. Therefore,
$$z(t)=\sqrt{z_0^2+c^2t^2}-z_0\eqno(2.4) $$
This law of motion, being manifestly relativistic, corresponds to
constant proper acceleration [12].\\
The charge velocity is given by
$$v=\frac{dz}{dt} =c^2t(z_0^2+c^2t^2)^{-1/2}.$$
Clearly, it tends to the light velocity in vacuum as $t\to\infty$.
The retarded times  $t'$ satisfy the following equation:
$$c_n(t-t')=[\rho^2+(z+z_0-\sqrt{z_0^2+c^2t'^2}\;)^2]^{1/2}\eqno(2.5)$$
It is convenient to introduce the dimensionless variables
$$\tilde t=ct/z_0,\quad \tilde z=z/z_0,\quad
\tilde\rho=\rho/z_0\eqno(2.6)$$
Then,
$$\alpha (\tilde t-\tilde t')=[\tilde{\rho}^2+(\tilde z+1-
\sqrt{1+\tilde t'^2}\;)^2]^{1/2},\eqno(2.7)$$
where $\alpha=c_n/c$ is the ratio of the light velocity in medium
to that  in vacuum.
In  order not to overload exposition we drop the tilde signs
$$\alpha(t-t')=[\rho^2+(z+1-\sqrt{1+t'^2}\;)^2]^{1/2}\eqno(2.8)$$
For the treated one-dimensional motion the denominators $R_i$ are
given by
$$R_i=\frac{z_0}{\alpha\sqrt{1+t_i^2}}[\alpha^2 (t-t_i)
\sqrt{1+t_i^2} - t_i(z+1-\sqrt{1+t_i^2}\;)]\eqno(2.9)$$

We consider the following two problems :\\
I. A  charged particle rests at the origin up to a moment $t'=0$.
After that it is accelerated in the positive direction of the $Z$ axis.\\
II. A charged particle decelerates  moving from $z=\infty$ to the origin.
After the moment $t'=0$ it rests there. \\
It is easy to check that the moving charge acquires
the light velocity $c_n$ at the moments
$t_l=\pm \alpha/\sqrt{1-\alpha^2}$ for the accelerated and
decelerated motion, resp.
The position of a charge at those moments is
$z_l=1/\sqrt{1-\alpha^2}-1 $.\\

It is our aim to investigate space-time distribution of  $EMF$
arising from such particle motions. For this, we should solve Eq.(2.8).
Taking its square we obtain the fourth degree algebraic equation
relative to $t'$. Solving it, we find space-time domains
where the $EMF$ exists.
It is just this way of finding  the $EMF$  which was adopted in [10].
It was shown in the same reference that there is another,
much simpler approach for recovering  $EMF$ singularities
(it was extensively used by Schott [6]).
We seek the zeros of the denominators $R_i$ entering into the definition of
the electromagnetic potentials (2.1). They are obtained from the equation
$$\alpha^2 (t-t')\sqrt{1+t'^2} - t'(z+1-\sqrt{1+t'^2}\;)=0\eqno(2.10)$$
We rewrite (2.8) in the form
$$\rho^2=\alpha^2(t-t')^2-(z+1-\sqrt{1+t'^2}\;)^2.\eqno(2.11)$$
Recovering $t'$ from (2.10) and substituting it into (2.11)
we find the surfaces $\rho(z,t)$ carrying the singularities
of the electromagnetic potentials. They are just shock waves
which we seek. It turns out that  $BS$ shock
waves (i.e., moving singularities arising from
the beginning or termination of a charge motion)
are not described by Eqs. (2.10) and (2.11). The physical
reason for this is that on these surfaces   $BS$ field strengths,
not potentials, are singular ([5]).
The simplified procedure mentioned above for recovering of moving $EMF$
singularities is to  find  solutions of (2.10)
and (2.11) and add to them "by hand " the positions of  $BS$
shock waves defined by the equation
$r=\alpha t,\quad r=\sqrt{\rho^2+z^2}$.
The equivalence of this approach to the complete solution of (2.8)
has been proved in [10] where the complete description
of the $EMF$ (not only its moving singularities as in the present approach)
of a moving charge was given.
It was shown there that  the electromagnetic potentials exhibited infinite
(for the Cherenkov and the shock waves under consideration) jumps when
one crosses the above singular surfaces.
Correspondingly, field strengths have
$\delta$-type singularities on these surfaces
while the space-time propagation of these
surfaces  describes the propagation of the radiated energy flux.
\section{Numerical results}
We consider the typical case when the ratio $\alpha$ of
the light velocity in medium to
that in vacuum is equal to 0.8.
\subsection{Accelerated motion}
For the first of the treated problems ( uniform acceleration of
the charge  resting at the origin up to a moment $t=0$) only positive retarded
times $t_i$ have a physical meaning (negative $t_i$ correspond to
the charge resting at the origin). The resulting configuration  of
the shock waves for the typical observation time $t=2$
is shown in Fig.1. We see  on it: \\
i) The Cherenkov shock wave $C_M$
having the form of the Cherenkov cone;\\
ii) The shock wave $C_L$ closing the Cherenkov cone  and
describing the shock wave  emitted from the point
$z_l=(1-\alpha^2)^{-1/2}-1$ at the moment $t_l=\alpha(1-\alpha^2)^{-1/2}$
when the velocity of a charge coincides with the light velocity
in  medium;\\
iii) The $BS$ shock wave $C_0$.\\
It turns out that the surface $C_L$ is approximated
with  good accuracy by the spherical surface
 $\rho^2+(z-z_l)^2=(t-t_l)^2$ (shown by the short-dash
curve $C$)
It should be noted that
only the part of $C$ coinciding with $C_L$ has physical meaning.

On the internal sides of the surfaces $C_L$ and $C_M$
electromagnetic potentials acquire  infinite values.
On the external side of $C_M$ lying outside
$C_0$ the magnetic vector potential is zero (as there are no solutions of
Eqs. (2.10),(2.11) there), while the electric scalar potential coincides
with that of the resting charge. On the external sides of $C_L$ and on
the part of the $C_M$ surface lying inside $C_0$ the electromagnetic
potentials have finite values (as bremsstrahlung has reached these space
regions).

In the negative $z$ semi-space the experimentalist will detect only the
$BS$ shock wave. In the positive $z$ semi-space, for the sufficiently
large times ($t>2\alpha/(1-\alpha^2)$) the observer close
to the $z$ axis will detect the Cherenkov shock wave $C_M$ first, the
$BS$ shock wave $C_0$ later and, finally, the shock wave $C_L$ originating from
the surpassing the medium light velocity.
For the observer more remoted from
the $z$ axis the $BS$ shock wave $C_0$ arrives first, then $C_M$ and finally
$C_L$ (Fig. 1).
For large distances from the $z$ axis the  observer  will see only
the $BS$ shock wave.

The positions of the shock waves for different observation times are
shown in Fig. 2. The dimension of the Cherenkov cone is zero
for $t\le t_l$ and continuously rises with time for $t>t_l$.
The physical reason for this is that the $C_L$ shock wave closing
the Cherenkov cone propagates with the light
velocity $c_n$, while the head part of the Cherenkov cone $C_M$
attached to the charged particle propagates with the velocity $v>c_n$.
It is seen that for small observation times ($t=2$ and $t=4$) the
$BS$ shock wave $C_0$ (pointed curve)
precedes  $C_M$. Later, $C_M$
overtakes (this happens at the moment $t=2\alpha/(1-\alpha^2)$)
and partly surpasses $BS$ shock wave $C_0$ ($t=8$).
However, the $C_L$ shock wave is always behind $C_0$ (as both of them
propagate with the velocity $c_n$, but $C_L$ is born
at the later moment $t=t_l$ ). The picture similar
to the $t=8$ case remains essentially the same for later times.

\subsection{Decelerated motion}
Now we turn to the second problem (uniform deceleration of the charged
particle along the positive $z$ semi-axis up to a moment $t=0$
after which it rests at the origin). In this case, only negative retarded
times $t_i$ have a physical meaning
(positive $t_i$ correspond to the charge resting at the origin).

For the observation time $t>0$ the resulting configuration of
the shock waves is shown in Fig. 3 where one sees the $BS$ shock wave $C_0$
arising from the termination of the charge motion
(at the moment $t=0$) and the blunt shock wave $C_M$ into which the $CSW$
transforms after the termination of the motion.
The head part of $C_M$  is described with  good accuracy  by the sphere
$\rho^2+(z-z_l)^2=(t+t_l)^2$
corresponding to the fictious shock wave $C$ emitted from the point
$z_l=(1-\alpha^2)^{-1/2}-1$ at the moment
$t_l=-\alpha(1-\alpha^2)^{-1/2}$ when the velocity of
the decelerated charge   coincides with
the medium light velocity.
Only part of $C$ coinciding with $C_M$ has  physical meaning.
The electromagnetic potentials vanish outside  $C_M$
(as no solutions exist there) and acquire  infinite values on
the internal part of $C_M$.
Therefore, the surface $C_M$ represents the shock wave. As a result,
for the decelerated motion after termination of
the particle motion ($t>0$)
one has the  shock wave $C_M$ detached from a moving charge
and the $BS$ shock wave $C_0$ arising
from the termination of the particle motion.  After the $C_0$ wave
reaches the observer, he will see the electrostatic field of a charge
at rest and bremsstrahlung from remoted parts of charge trajectory.

The positions of  shock waves for different  times are shown
in Fig. 4 where  one sees  how the acute  $CSW$,
attached to the moving charge ( $t=-2$), transforms into
the blunt shock wave detached from it ($t=8$). Pointed curves
mean  the $BS$ shock waves described by the equation $r=\alpha t$.

For the decelerated motion and $t<0$ (i.e., before termination of
the charge motion)  physical solutions  exist only inside
the Cherenkov cone $C_M$ ( $t=-2$ on Fig. 4).
On the internal boundary of the Cherenkov cone the electromagnetic
potentials acquire  infinite values.
On their external boundaries the electromagnetic potentials are zero
(as no solutions exist there).
When the charge velocity coincides with $c_n$
the $CSW$ leaves the charge and transforms into the $C_M$
shock wave which propagates  with the velocity
$c_n$ ($t=2,\;4$ and $8$ on Fig. 4).
As it has been mentioned, the blunt head parts
of these waves are approximated with a good accuracy by the surface
$\rho^2+(z-z_l)^2=(t+t_l)^2$  corresponding to the fictious shock
wave emitted at the moment when the charge velocity coincides with
the light velocity in the medium. \\
In the negative $z$ semi-space the experimentalist  will detect
the blunt shock wave first and $BS$ shock wave (shock dash curve) later.\\
In the positive $z$ semi-space,
for the observation point close to the $z$ axis the observer will see
the $CSW$ first and $BS$ shock wave later.
For larger distances from the $z$ axis he will see at first
the blunt shock $C_M$ into which the $CSW$ degenerates after the
termination of the charge motion and the $BS$ shock wave  later (Fig. 4).

It should be mentioned  about the
continuous radiation which reaches the observer between the arrival of
the above shock waves and about the continuous radiation and the
electrostatic field of a charge at rest after the arrival of the last
shock wave. They are  easily restored from the complete exposition
presented in [10] for the $z=at^2$ motion law.

\section{Conclusion}
We have investigated the space-time distribution of
the electromagnetic field arising from the accelerated manifestly
relativistic charge motion.
This motion is maintained by the constant electric field.
Probably, this field is easier to create in gases (than in solids
in which the screening effects are essential)
where the Vavilov-Cherenkov effect is also observed.
We have confirmed the intuitive predictions made by Tyapkin ([2]) and
Zrelov et al. ([3]) concerning the existence of the new shock wave
(in addition to the Cherenkov and bremsstrahlung shock waves)
arising when the charge velocity coincides with the  light velocity
in medium. For the accelerated motion, this shock wave forms
indivisible unity with Cherenkov's shock wave. It closes the
Vavilov-Cherenkov radiation cone and propagates with the light velocity
in the medium. For the decelerated motion, the  above
shock wave detaches from a moving charge when its velocity
coincides with the light velocity in medium.\\
The quantitative conclusions made in [10]
for a less realistic external electric field maintaining the accelerated charge
motion are also confirmed. We have specified under what conditions and in which
 space-time regions the above-mentioned new shock waves do exist.
It would be interesting to observe these  shock waves experimentally.

\eject
\begin{figure}[h]
\vspace*{-15mm}
\begin{center}
\includegraphics[angle=-90,width=90mm]{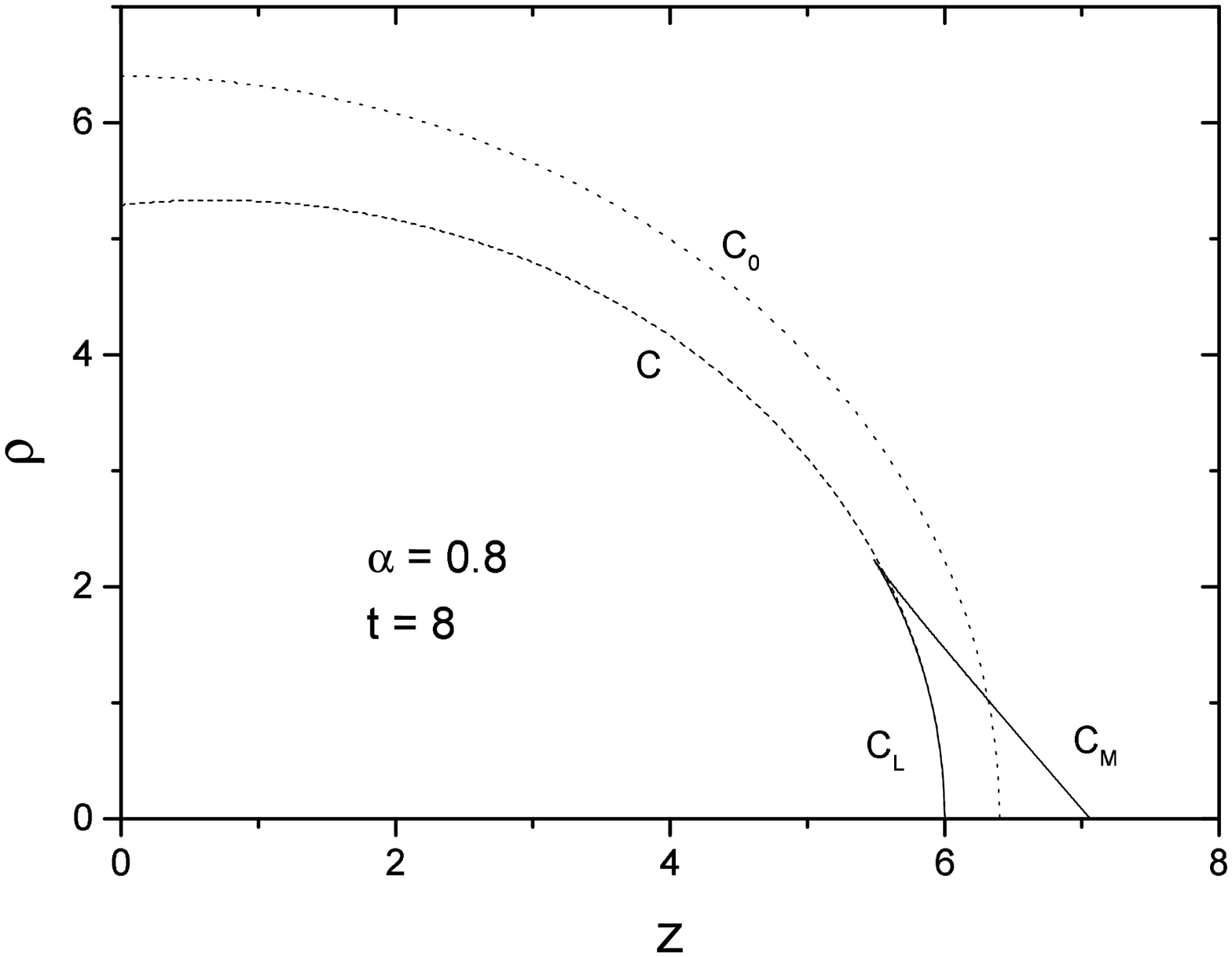}
\end{center}
\vspace*{-5mm}
\caption{\small Typical distribution of the shock waves
emitted by  the accelerated charge.
$C_M$ is the Cherenkov shock wave,
$C_L$ is the shock wave emitted from the point
$z_l= (1-\alpha^2)^{-1/2}$
at the moment $t_l=\alpha(1-\alpha^2)^{-1/2}$
when the charge velocity
coincides with the medium light velocity.
Part of it is described with  good accuracy
by the fictious spherical
surface $C$ ($\rho^2+(z-z_l)^2=(t-t_l)^2$). $C_0$ is the
bremsstrahlung shock wave originating from
the beginning ( at the moment $t=0$) of the charge motion.}
\end{figure}
\begin{figure}[h]
%\vspace*{-8mm}
\begin{center}
%\hspace*{-20mm}
\includegraphics[angle=-90,width=90mm]{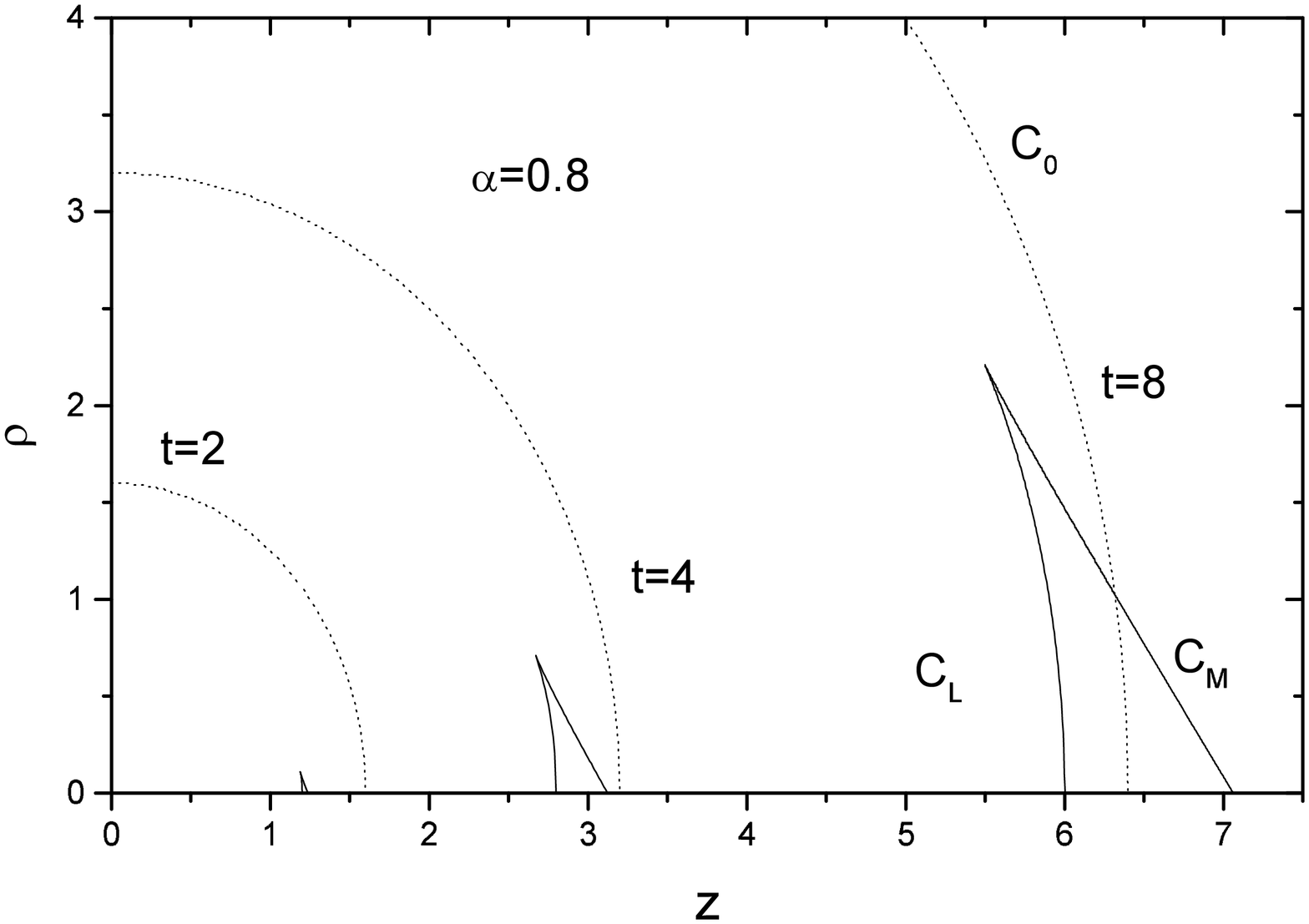}
%\vspace*{-2mm}
\caption{\small Time  evolution of shock waves
emitted by  the accelerated
charge. $C_M$ and $C_L$ are respectively the usual
Cherenkov shock wave
and the shock wave arising at the moment when the charge velocity
coincides with the medium light velocity.
Pointed curves are bremsstrahlung shock waves.}
\end{center}
\vspace*{-45mm}
\end{figure}
\newpage
\begin{figure}[h]
\vspace*{-15mm}
\begin{center}
\includegraphics[angle=-90,width=90mm]{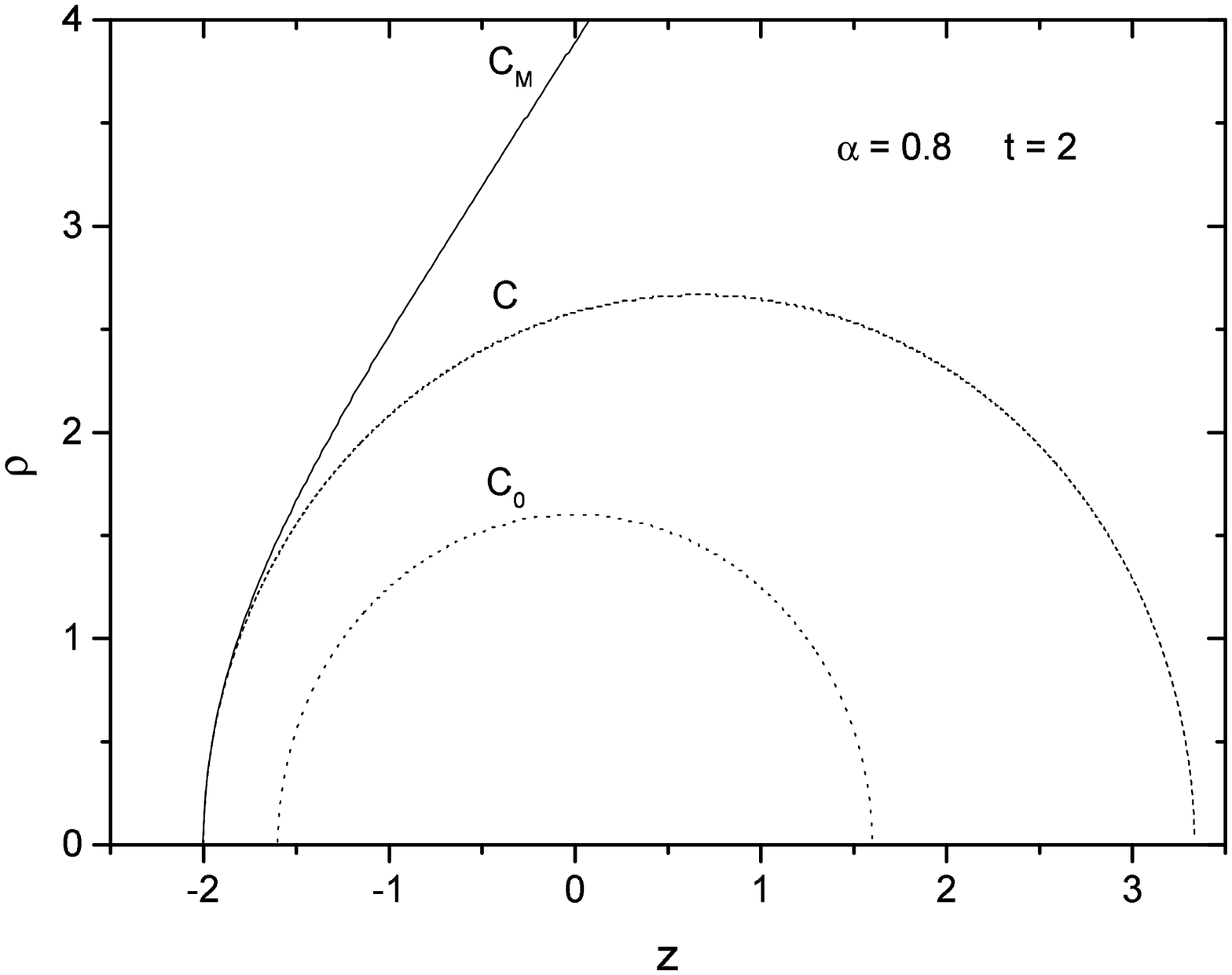}
\end{center}
\vspace*{-5mm}
\caption{\small Space  distribution of the shock waves
produced by the
decelerated charge  in the uniform electric field.
$C_M$ is the blunt shock wave into which the $CSW$ transforms
after the moment when the charge velocity coincides with the medium
light velocity. Part of it is approximated with  good accuracy
by the fictious spherical surface $C$.
$C_0$ is the bremsstrahlung shock wave originating from
the termination of the charge motion at the moment $t=0$.}
\end{figure}
\begin{figure}[h]
%\vspace*{-8mm}
\begin{center}
\includegraphics[angle=-90,width=90mm]{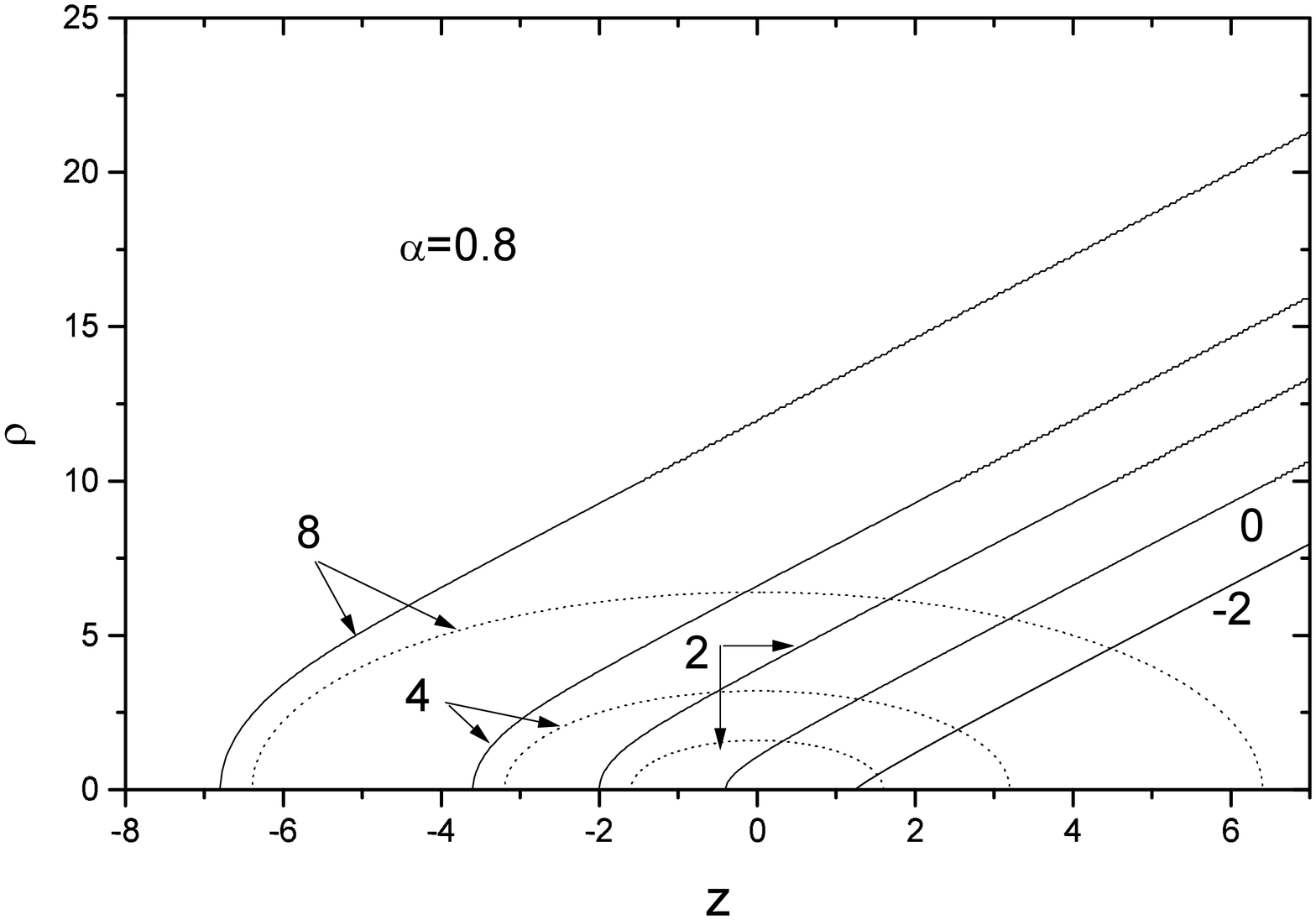}
\vspace*{-2mm}
\caption{\small The continuous transformation of
the acute Cherenkov shock wave  attached to a moving charge ($t=-2$)
into the blunt shock wave detached from a charge ($t=8$)
for the decelerated motion.
The numbers at the curves  mean the observation times.
Pointed curves are bremsstrahlung shock waves.
 Charge motion is terminated at the moment $t=0$.}
\end{center}
\vspace*{-45mm}
\end{figure}


\newpage
\begin{thebibliography}{99}
\bibitem{1}
I.M. Frank, Vavilov-Cherenkov Radiation. Theoretical Aspects
(Nauka, Moscow, 1988).
\bibitem{2}
A.A. Tyapkin, JINR Rapid Communications, Dubna, \textbf{3}, 26 (1993).
\bibitem{3}
V.P. Zrelov, J. Ruzicka and A.A. Tyapkin,
JINR Rapid Communications, Dubna, \textbf{1[87]-98}, 10 (1998).
\bibitem{4}
G.M. Volkoff, Amer. J. Phys., \textbf{31}, 601 (1963).
\bibitem{5}
G.N. Afanasiev, Kh. Beshtoev and Yu.P. Stepanovsky,
Helv. Phys. Acta,  \textbf{69}, 111 (1996);\\
G.N. Afanasiev, V.G. Kartavenko and Yu.P. Stepanovsky,
J. Phys. D:  Appl. Phys.  \textbf{32}, 2029 (1999).
\bibitem{6}
G.A. Schott, Electromagnetic Radiation
(Cambridge Univ. Press, Cambridge, 1912).
\bibitem{7}
I.E. Tamm, J. Phys. USSR  \textbf{1}, No 5-6, 439 (1939).
\bibitem{8}
V.P. Zrelov and J. Ruzicka, Chech. J. Phys. B,  \textbf{39}, 368 (1989).
\bibitem{9}
V.P. Zrelov and J. Ruzicka, Chech. J. Phys.,  \textbf{42}, 45 (1992).
\bibitem{10}
G.N. Afanasiev, S.M. Eliseev and Yu.P. Stepanovsky,
Proc. R. Soc. Lond., ser. A (Mathematical, Physical and Engineering Sciences),
 \textbf{454}, 1049 (1998).
\bibitem{11}
G.N. Afanasiev and V.G. Kartavenko,
J. Phys. D:  Appl. Phys.,  \textbf{31}, 2760 (1998);\\
G.N. Afanasiev, V.G. Kartavenko and E.N. Magar,
Physica B  \textbf{269}, 95 (1999).
\bibitem{12}
L.D. Landau and E.M. Lifshitz,
The Classical Theory of Fields (Pergamon, New York, 1962).
\end{thebibliography}
\end{document}